\documentclass[aps,pra,showpacs,nontitlepage, twocolumn]{revtex4}
    \usepackage{amsmath}
    \usepackage{amssymb}
\usepackage{bm}

\usepackage{natbib}
\usepackage{hyperref}
\usepackage{graphicx}

\renewcommand\Re{\operatorname{Re}}
\renewcommand\Im{\operatorname{Im}}
\renewcommand{\vec}[1]{\mathbf{#1}}

\newcommand{\epl}{\ensuremath{\varepsilon_\parallel}}
\newcommand{\epr}{\ensuremath{\varepsilon_\perp}}

\newcommand{\sepl}{\ensuremath{\sqrt{\varepsilon_\parallel}}}
\newcommand{\sepr}{\ensuremath{\sqrt{\varepsilon_\perp}}}

\newcommand{\eps}{\ensuremath{\hat \varepsilon}}
\newcommand{\epsn}{\ensuremath{\hat \epsilon}}

\newcommand{\rr}{\ensuremath{\vec r}}
\newcommand{\xx}{\ensuremath{\vec {\hat x}}}
\newcommand{\yy}{\ensuremath{\vec {\hat y}}}
\newcommand{\zz}{\ensuremath{\vec {\hat z}}}

\newcommand{\GF}{\ensuremath{{\hat G}}}

\newcommand{\rc}{\ensuremath{\rr\times\zz}}

\newcommand{\eikro}{\ensuremath{e^{ikr_o}}}
\newcommand{\eikre}{\ensuremath{e^{ikr_e}}}

\begin{document}

\title{Green function for hyperbolic media}

\author
{
Andrey S. Potemkin,${}^1$ Alexander N. Poddubny,${}^{1,2}$ Pavel A. Belov,${}^{1,3}$ and Yuri S. Kivshar,${}^{1,4}$
}

\address
{
$^{1}$Department of Photonics and Optoinformatics, National University of Information Technology, Mechanics and Optics (ITMO), St.~Petersburg 197101, Russia\\
$^{2}$Ioffe Physical-Technical Institute of the Russian Academy of Sciences,  St.~Petersburg 194021, Russia\\
$^{3}$School of Electronic Engineering and Computer Science, Queen Mary University of London, Mile End Road, London E1 4NS, UK\\
$^{4}$Nonlinear Physics Center, Research School of Physics and Engineering, Australian National University, Canberra ACT 0200, Australia
}
\begin{abstract}
We revisit the problem of the electromagnetic Green function for homogeneous hyperbolic media, where longitudinal and transverse components of the 
dielectric permittivity tensor have different signs. We analyze the dipole emission patterns for both dipole orientations with respect to the symmetry axis 
and for different signs of dielectric constants, and show that the emission pattern is highly anisotropic and has a characteristic cross-like shape: 
the waves are propagating within a certain cone and are evanescent outside this cone. We demonstrate the coexistence of the cone-like pattern due to emission 
of the extraordinary TM-polarized waves and elliptical pattern due to emission of ordinary TE-polarized waves.
We find a singular complex term in the Green function, proportional to the $\delta-$function and governing the photonic density of states and Purcell effect in hyperbolic media.
\end{abstract}

\maketitle

\section{Introduction}

Hyperbolic medium is a particular case of uniaxial anisotropic dielectric medium where the main values of the permittivity tensor $\hat{\varepsilon}$ have opposite 
signs~\cite{Felsen1972}, and the isofrequency surface of extraordinary waves is a hyperboloid. The recent attention of researchers to the hyperbolic metamaterials 
stems from their unique optical properties allowing negative refraction, hyperlensing and cloaking phenomena~\cite{Smith2003,Smith2004,Cai2008}. They are also very 
promising  for quantum nanophotonics~\cite{Krishnamoorthy2012,shalaev2011}, because the density of states, determined by the hyperboloid area, is divergent. 
This means an infinite spontaneous decay rate of a quantum emitter embedded in a hyperbolic medium, i.e. infinite Purcell effect~\cite{Purcell}. 
In the realistic case, the radiation rate is limited by certain cutoffs in the wavevector space~\cite{Poddubny2011,sipe2011,Iorsh2012,Yan2012arXiv},
 however, experimental observation of radiative enhancement is still possible~\cite{Krishnamoorthy2012,Kim2012}.

\begin{figure}[t]
\centering\includegraphics[width=0.48\textwidth]{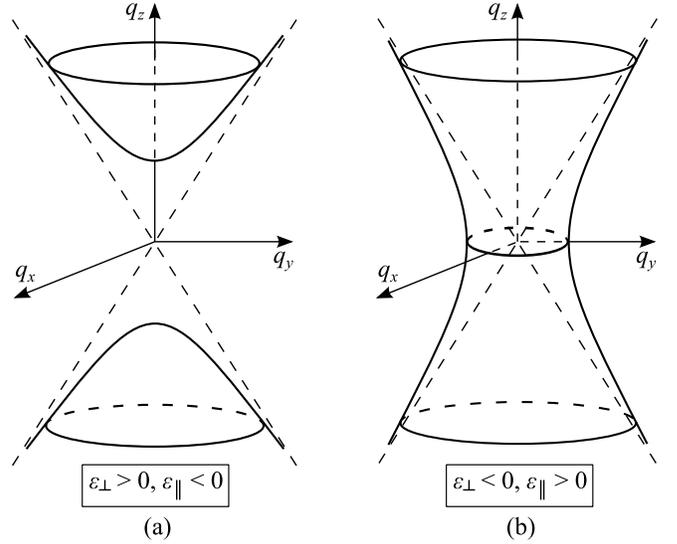}
\caption{Schematic illustration of the isofrequency surfaces in the wavevectors space for the hyperbolic medium with
$\varepsilon_{\parallel}<0$,
$\varepsilon_{\perp}>0$ (a) and with $\varepsilon_{\parallel}<0$, $\varepsilon_{\perp}>0$ (b).
}\label{fig:cones}
\end{figure}

A general hyperbolic medium can be characterized by a dielectric permittivity tensor
\begin{equation}
\hat\varepsilon =\left(\begin{array}{ccc}
                    \epr & 0 & 0\\0 & \epr & 0\\ 0 & 0 & \epl
                  \end{array}\right)\:,\quad
 \Re{\epr}\Re{\epl} < 0\:. \label{eq:EpsTensor}
\end{equation}
Resulting isofrequency surface of extraordinary waves is a hyperboloid,
\begin{equation}
\frac{q_x^2+q_y^2}{\varepsilon_{\parallel}} +\frac{q_z^2}{\varepsilon_{\perp}}=\left(\frac{\omega}{c}\right)^2\:.\label{eq:hyp}
\end{equation}
Two different types of hyperboloids are possible for $\varepsilon_{\parallel}<0$,
$\varepsilon_{\perp}>0$ (see Fig.~\ref{fig:cones}a) and for $\varepsilon_{\parallel}>0$, $\varepsilon_{\perp}<0$ (see Fig.~\ref{fig:cones}b).

Hyperbolic medium can be realized in several ways. First  realization has been reported
for magnetized plasma in microwave spectral range~\cite{Fisher1969}.
Under a very strong static magnetic field applied along $z$ axis the plasma is described by the permittivity with~\cite{Fisher1969, Zhang2011}
$$
\epr = 1,\quad \epl = 1-\frac{\omega_p^2}{\omega^2}.
$$
 Thus, the frequencies $\omega$ below the plasma frequency $\omega_p$ correspond to the hyperbolic regime with $\epl<0$.

The interest to the hyperbolic medium is now revived and rapidly increases due to its successful realization  using artificial photonic structures, metamaterials~\cite{Yao2008,narimanov2009b,narimanov2010b,Krishnamoorthy2012,Cortes2012arXiv}.
In particular, the layered structure composed of alternating dielectric and metallic slabs is described by effective permittivities
$$
\epr = \frac{\varepsilon_1 d_1+\varepsilon_2 d_2}{d_1+d_2},\quad \epl=\left(\frac{\varepsilon_1^{-1}d_1 + \varepsilon_2^{-1}d_2}{d_1+d_2}\right)^{-1},
$$ where
$\varepsilon_1$, $\varepsilon_2$, $d_1$, $d_2$ are dielectric constants and thicknesses of slabs, respectively. Since dielectric constants of the metals in optical frequency range are negative, it is possible to adjust thicknesses of slabs in order to obtain either $\epr<0$ or $\epl<0$~\cite{orlov2011,chebykin2011}. Realizations of hyperbolic regime have been also reported for metamaterials based on nanorod arrays~\cite{narimanov2009b,Wurtz2008,simovski2012} and for graphite~\cite{Sun2011}.

The ongoing  studies of hyperbolic medium raise the demand for the general theoretical formalism.
 The most rigorous  description of the optical properties of arbitrary photonic structure is provided by its tensor Green function~\cite{Welsch2006},
determined from
\begin{equation}
(\nabla\times\nabla\times\hat{I} - k^2\eps)\GF(\bm r)=4\pi k^2{\hat I}\delta(\rr)\label{eq:GreenEquation}\:,
\end{equation}
where $\hat I$ is the unit tensor and $k=\omega/c$.
  Green function for medium with uniaxial permittivity tensor  is presented in a number of works \cite{Felsen1972,Clemmow1963,Chen1983,weiglhofer:1095,Weiglhofer1989,Weiglhofer1990,Lindell1992,cottis1999, Savchenko2005,Sautbekov2008}. Most general form of the result is given by Chen in Ref.~\cite{Chen1983} in dyadic form and by Savchenko in Ref.~\cite{Savchenko2005} in Cartesian form. However, despite the impressive amount of the researchers done, we are not aware of any comprehensive study of the Green function in hyperbolic case. Moreover, all existing works except \cite{Weiglhofer1990, cottis1999}  neglect the singular term in the Green function, which, as will be shown below, is crucial for the description of the photonic density of states and Purcell enhancement in hyperbolic regime. Here we present a general theory of the Green function and dipole emission in hyperbolic medium. We analyze both types of hyperbolic medium, illustrated on Fig.~\ref{fig:cones},
and both axial and transverse dipole orientations.

The rest of the paper is organized as follows. Sec.~\ref{sec:Green} outlines the Green function calculation. Singular term in the Green function is discussed in Sec.~\ref{sec:Sing}. Sec.~\ref{sec:Pattern} presents the calculated emission patterns of point dipole. Main paper results are summarized in Sec.~\ref{sec:Concl}.
Auxiliary expressions are presented in Appendices.

\section{Green function calculation}\label{sec:Green}
Green function \eqref{eq:GreenEquation} may be calculated either in real space via the methods of operators~\cite{Lindell1992} or in wavevector space via Fourier analysis~\cite{Chen1983}.
The Fourier representation of $\GF(\rr)$, defined from
\begin{equation}
 \GF(\rr)=\int \frac{d^3q}{(2\pi)^3}e^{i\bm q\bm r}\GF(\bm q)\:,
\end{equation}
is given by
\begin{multline}
\GF(\vec q)=4\pi k^2\left\lbrace\left(\epr\epl\eps-\frac{\vec q\otimes\vec q}{k^2}\right)\frac{1}{q_\parallel^2\epl+q_\perp^2\epr-k^2\epr\epl}\right.\\
+\left.\frac{(\vec q\times\hat{\bf z})\otimes(\vec q\times\hat{\bf z})}{q_\perp^2}\left(\frac{1}{q^2-k^2\epr}\right.\right.\\\left.\left.-\frac{\epl}{q_\parallel^2\epl+q_\perp^2\epr-k^2\epr\epl}\right)\right\rbrace.\label{eq:Gq}
\
 \end{multline}
Here $q_\parallel\equiv q_z$  and $q_\perp^2=q_x^2+q_y^2$. The symbol $\otimes$ denotes the direct product, $[\bm a\otimes \bm b]_{\alpha\beta}\equiv a_{\alpha}b_\beta$.
The poles in Eq.~\eqref{eq:Gq} determine the dispersion equations  of the electromagnetic waves. Two poles correspond to extraordinary (TM) waves, with magnetic field $\bf H$ perpendicular to $z$ axis,
 and to ordinary (TE)  waves, with $\bf E\cdot\hat{\bf z}=0$ and $q=k\sqrt{\epr}$.

Both direct and reciprocal space methods give the same following result for the Green function:
\begin{align}
 \GF(\rr)&=\frac{1}{\sepr}\left\lbrace\left(k^2\epsn+\nabla\otimes\nabla\right)\frac{\eikre}{r_e}\right.\nonumber\\
&+k^2\left(\epr\frac{\eikro}{r_o}-\epl\frac{\eikre}{r_e}\right)\frac{(\rc)\otimes(\rc)}{(\rc)^2}\label{eq:SimpleG}\\
&-ik\left.\frac{\eikro-\eikre}{(\rc)^2}\left(\vec{\hat I} - \zz\otimes\zz - \frac{2(\rc)\otimes(\rc)}{(\rc)^2}\right)\right\rbrace\nonumber,
\end{align}
where
\begin{gather*}
\epsn\equiv \epr\epl\eps^{-1} =\epl(\vec{\hat I}-\zz\otimes\zz)+\epr\zz\otimes\zz,\\
r_e =\sqrt{\rr(\epsn\rr)}=\sqrt{\epl(x^2+y^2)+\epr z^2},\quad r_o =\sepr r.
\end{gather*}
This result was obtained without any assumptions for signs of real of parts $\epr$ and $\epl$ and is applicable to case of hyperbolic medium. Imaginary parts of $\epr$ and $\epl$ and of the square roots in expressions for $r_e$ and $r_o$ should be positive (we assume the time dependence $e^{-i\omega t}$).

Eq.~\eqref{eq:SimpleG} is not the final result for the Green function yet, because the
 derivative in the term ${\nabla\otimes\nabla \eikre/r_e}$ is not evaluated. Some authors~\cite{Lindell1992,Sautbekov2008}  leave the expression for $\GF(\rr)$ in the form \eqref{eq:SimpleG} without taking this derivatives. However, this term produces singularity in $\GF(\rr)$. The nature of this singularity is the same for the well-known identity
\begin{equation*}
 \Delta \frac{1}{r}=-4\pi\delta(\bm r)\:.
\end{equation*}
%

Calculating the derivative we obtain the  final expression of singular Green function:
\begin{align}
\GF(\rr)&=\GF_{\rm sing}(\rr)+\frac{1}{\sepr}\Biggl\{
k^2\frac{\eikre}{r_e}\left(1-\frac{1}{ikr_e}-\frac{1}{k^2 r_e^2}\right)\epsn\nonumber\\
&-k^2\frac{\eikre}{r_e^3}\left(1-\frac{3}{ikr_e}-\frac{3}{r_e^2}\right)(\epsn\rr)\otimes(\epsn\rr)\label{eq:FinalG}\\
&+k^2\left(\epr\frac{\eikro}{r_o}-\epl\frac{\eikre}{r_e}\right)\frac{(\rc)\otimes(\rc)}{(\rc)^2}\nonumber\\
&-ik\frac{\eikro-\eikre}{(\rc)^2}\left(\vec{\hat I} - \zz\otimes\zz - \frac{2(\rc)\otimes(\rc)}{(\rc)^2}\right)\Biggr\}\nonumber\:,
\end{align}
where $\GF_{\rm sing}(\rr)$ stands for the singular contribution
\begin{multline}
 \GF_{\rm sing}(\rr)=\left[(\nabla\otimes\nabla)\frac1{\sepr r_e}\right]_{\rm sing}\label{eq:Gsing1}\\\equiv -4\pi\int \frac{d^3q}{(2\pi)^3}e^{i \vec q\vec r}\frac{ \vec q\otimes\vec q}{q_\parallel^2\epl+q_\perp^2\epr}\:.
\end{multline}
In general case we failed to obtain a closed answer for $\GF_{\rm sing}$ via ordinary functions and $\delta$-function. Eq.~\eqref{eq:Gsing1} should be understood instead only as generalized function, i.e., only its convolutions with ordinary test functions are relevant~\cite{Gelfand}. Singular Green  function is essential in the hyperbolic case since it solely accounts for the diverging Purcell factor.
Eqs.~\eqref{eq:FinalG},\eqref{eq:Gsing1} are the central result of this work. It is valid for both hyperbolic and elliptic media with arbitrary signs of real parts of dielectric constants.

\section{Singular term of Green function}\label{sec:Sing}
\begin{figure*}[t]
\begin{center}
\includegraphics[width=\textwidth]{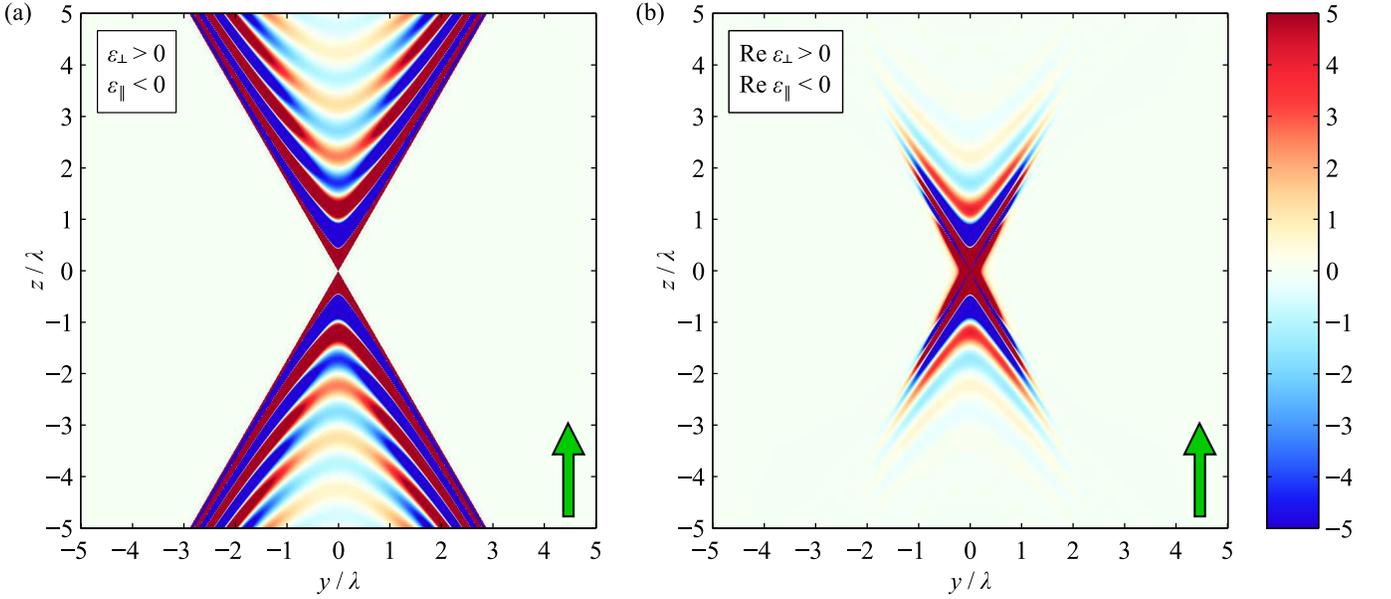}
\end{center}
\caption{Dipole field $\lambda^3 E_z(0, y, z)$ in hyperbolic medium with $\epr = 1$, $\epl = -3$ (a) and with $\epr = 1 + 0.2i$, $\epl = -3 + 0.2i$ (with losses) (b). Dipole moment is parallel to the axis of anisotropy~\zz.}\label{fig:Epar1}
\end{figure*}
Singular term in the Green function should be treated with extra care~\cite{Yaghjian1980,Tai1981,Wang1982}.
Functional~\eqref{eq:Gsing1} may be explicitly written out in only isotropic medium
 with $\epl=\epr\equiv\varepsilon$~\cite{Franklin2010}:
\begin{equation}\label{eq:RightDelta}
 \GF_{\rm sing}(\rr)=-\frac{4\pi}{\varepsilon} \frac{\rr\otimes \rr}{r^2}\delta(\bm r)\:.
\end{equation}
In general anisotropic case we have not found a closed form answer. Previous expression for this functional in anisotropic medium,
obtain by Weiglhofer~\cite{Weiglhofer1990, Weiglhofer1989},
\begin{equation}
-\frac{4\pi}{3}\eps^{-1}\delta(\rr)\label{eq:Weiglhofer}
\end{equation}
is obviously wrong since it has been obtained as a naive generalization of the  result \cite{Frahm1983}
\begin{equation}
-\frac{4\pi}{3 \varepsilon}\delta(\rr)\label{eq:ConvDelta}
\end{equation}
for the isotropic medium.
Despite the fact that functional~\eqref{eq:ConvDelta} is frequently used  \cite{Frahm1983},  it  gives different results than Eq.~\eqref{eq:RightDelta}
when applied to functions, non-analytical in the point $\rr=0$, such as $\rr\otimes \rr/r^2$.
 This discrepancy is discussed in~Ref.~\cite{Franklin2010} in details. Since angular averaging of $\rr\otimes\rr$ equals to $1/3$,
  both functionals~\eqref{eq:ConvDelta} and \eqref{eq:RightDelta} may give the same result for some test functions. However, straightforward generalization of
Eq.~\eqref{eq:ConvDelta} to Eq.~\eqref{eq:Weiglhofer} is not valid  since not all rules of normal calculus may be applied to generalized functions~\cite{Gelfand}.

Explicit form of the functional \eqref{eq:Gsing1} in anisotropic case can be  obtained
within the space of the test functions, analytical in the point $\bm r=0$:
\begin{multline}\label{eq:Gsing2}
 \hat{\tilde G}_{\rm sing}(\bm r)=\frac{2\pi \delta(\bm r)}{\epl-\epr}\Biggl[\hat{\bm x}\otimes \hat{\bm x}+\hat{\bm y}\otimes\hat{\bm y}-2\hat{\bm z}\otimes\hat{\bm z}\\
+\arctan(\sqrt{\epl/\epr-1})\left(\frac{2}{\sqrt{\epl/\epr-1}}\hat{\bm z}\otimes\hat{\bm z}\right.\\-\left.\frac{\epl}{\epr\sqrt{\epl/\epr-1}}(\hat{\bm x}\otimes \hat{\bm x}+\hat{\bm y}\otimes\hat{\bm y})\right)\Biggr]\:.
\end{multline}
The equivalence between Eq.~\eqref{eq:Gsing2} and Eq.~\eqref{eq:Gsing1} for test functions, analytical in the point $r=0$ is directly shown  via real space integration in spherical coordinates. Let us demonstrate it for  $zz$ component:
\begin{multline}
 \int\limits_{r<R}d^3r [G_{\rm sing}]_{zz}(\bm r)=\int\limits_{r<R}d^3r
\frac{\partial^2}{\partial z^2}\frac1{\sepr r_e}\\=\oint\limits_{r=R} d\bm{S}\cdot \zz
\frac{\partial}{\partial z}\frac1{\sepr r_e}\\=-2\pi\sqrt{\epr} \int_0^\pi
d\theta \frac{\sin \theta\cos^2\theta}{(\epl\sin^2\theta+\epr\cos^2\theta)^{3/2}}\:.
\end{multline}
Performing the integral over $\theta$ we obtain expression, equal to
$\int d^3r[\tilde G_{\rm sing}]_{zz}(\bm r)$ which finalizes the proof.  Eq.~\eqref{eq:Gsing2} also follows from  the  wavevector-space representation in Eq.~\eqref{eq:Gsing1}.
In the limit $\epr=\epl$ Eq.~\eqref{eq:Gsing2} reduces to Eq.~\eqref{eq:ConvDelta}.
Obviously, functional~\eqref{eq:Gsing2} is far more complex than (wrong) Eq.~\eqref{eq:Weiglhofer}.
 Still,  it is valid for narrower set of test functions than Eqs.~\eqref{eq:Gsing1},\eqref{eq:RightDelta}.

One may think that this singular terms are of purely mathematical interest.
Counterintuitively,  they may control such observable quantities, as  photonic density of states and Purcell factor.
In particular, in the case of lossless hyperbolic medium ($\epl\epr<0$, $\Im \epl,\epr\to+0$)
 the singular part of the Green function acquires non-zero imaginary part,
\begin{multline}
\label{eq:DOS}
 \Im \hat {\tilde G}_{\rm sing}(\bm r)=\frac{2\pi^2\sqrt{|\epr|}}{(|\epr|+|\epl|)^{3/2}}\delta(\rr)\\\times
\left[\zz\otimes\zz +\frac{|\epl|}{|\epr|}(\xx\otimes \xx+\yy\otimes\yy)\right]\:.
\end{multline}
The singulary in Eq.~\eqref{eq:DOS} is a direct consequence of the infinite density of TM modes Eq.~\eqref{eq:hyp}. Green function where this singularity is omitted has wrong analytical properties and can not be used for nanophotonic applications such as Purcell factor calculation.
Generally, the Purcell factor of the embedded light emitter oriented along $\hat{\bf n}$ direction can be found\cite{Novotny2006} as
\begin{equation}\label{eq:Purc1}
 f=\frac{3}{2k^3}\Im \hat{\bf n}\cdot \hat G(0)\hat{\bf n}\:.
\end{equation}
Due to the singularity in the local density of states Eq.~\eqref{eq:Purc1} provides diverging Purcell factor.
As has been indicated in
our previous work \cite{Poddubny2011}, this divergence is smeared out for finite size emitter,
characterized with spatial distribution $\Phi(\bm r)$, normalized as $\int d^3r \Phi(\bm r)=1$.
In case of the semiconductor quantum  dot $\Phi(\bm r)$ is proportional to the exciton envelope function~\cite{Ivchenko2005}. For distributed source  Eq.~\eqref{eq:Purc1} is replaced by
\begin{equation}\label{eq:Purc2}
 f=\frac{3}{2k^3}\hat{\bf n}\cdot \int d^3rd^3r'\Phi(\bm r)\Phi(\bm r')\hat G(\bm r-\bm r')
\hat{\bf n}\:.
 \end{equation}
 In the case of isotropic source distribution, $\Phi(\bm r)\equiv \Phi(r)$,  the Purcell factor is readily evaluated using \eqref{eq:DOS} and is proportional to the cube of the ratio of the wavelength and the size of the source.

\begin{figure*}[t]
\begin{center}
\includegraphics[width=\textwidth]{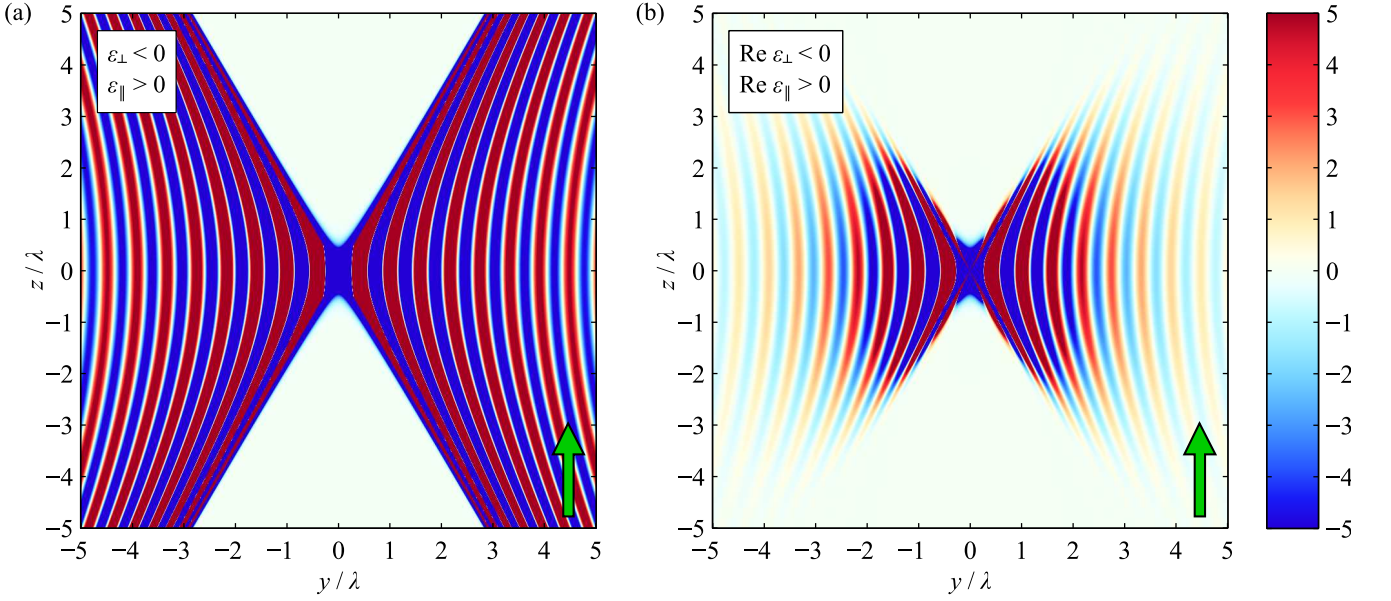}
\end{center}
\caption{Dipole field $\lambda^3 E_z(0, y, z)$ in hyperbolic medium with $\epr = -1$, $\epl = 3$ (a) and with $\epr = -1 + 0.2i$, $\epl = 3 + 0.2i$ (with losses) (b). Dipole moment is parallel to the axis of anisotropy~\zz.}\label{fig:Epar2}
\end{figure*}

\section{Dipole emission pattern}\label{sec:Pattern}
Green function \eqref{eq:GreenEquation} allows to find electric field for arbitrary distribution of polarization $\vec P(\vec r)$,
\begin{equation}
 \vec E(\vec r)=\int d^3r' \hat G(\vec r-\vec r')\vec P(\vec r')\:.
\end{equation}
For the point dipole $\bm p$ one has $\bf P(\bf r)=\bf p\delta(\bf r)$ and  the electric field is given by
\begin{equation}
\vec E(\rr)=\GF(\rr)\cdot \vec p\label{eq:SimpleE}\:.
\end{equation}
In the following subsections we consider  two principal cases of orientation of dipole moment $\vec p$ with respect to the anisotropy axis~$\zz$.

It should be noted that the regular part of the dipole field may be easily obtained without knowledge of the dyadic Green function~\eqref{eq:FinalG}.
Electromagnetic field in the uniaxial medium can be decomposed into two parts: TM-field, where $\vec H\cdot\zz$ is zero, and TE-field, where $\vec E\cdot\zz$ is zero~\cite{Clemmow1963}. Solutions for these fields in hyperbolic medium can be found separately via corresponding solutions in vacuum using of anisotropic scaling method introduced by Clemmow in Ref.~\cite{Clemmow1963a}. The key point of this method consists in appropriate scaling of space as well as fields and polarizations from Maxwell's equations in vacuum to obtain expressions for fields and currents in anisotropic medium. Application of Clemmow's method for hyperbolic medium gives the following scaling rules:\\
(i) TE polarization
\begin{align}
\vec E(\rr) &= \epsn^{1/2}\vec E_0(\epsn^{1/2}\rr)\label{ETMScale},\\
\vec H(\rr) &= \sqrt{\epr\epl}\vec H_0(\epsn^{1/2}\rr)\label{HTMScale},\\
\vec P(\rr) &= \sqrt{\epr\epl}\eps^{1/2}\vec P_0(\epsn^{1/2}\rr)\label{JTMScale},
\end{align}
(ii) TM polarization
\begin{align}
\vec E(\rr) &= \vec E_0(\sepr\rr)\label{ETEScale},\\
\vec H(\rr) &= \sepr\vec H_0(\sepr\rr)\label{HTEScale},\\
\vec P(\rr) &= \epr\vec P_0(\sepr\rr)\label{JTEScale},
\end{align}
where $\bm E_0, \bm H_0$, $\bm P_0$ are vacuum solutions and
\begin{gather*}
\eps^{1/2}=\sepr(\vec{\hat I}-\zz\otimes\zz)+\sepl\zz\otimes\zz,\\
\epsn^{1/2}=\sepl(\vec{\hat I}-\zz\otimes\zz)+\sepr\zz\otimes\zz.
\end{gather*}
Decomposition of the dipole polarization $\vec P$ into the TM/TE parts is described in~\cite{Lindell1988, Lindell1990}.

\begin{figure*}[t]
\begin{center}
\includegraphics[width=\textwidth]{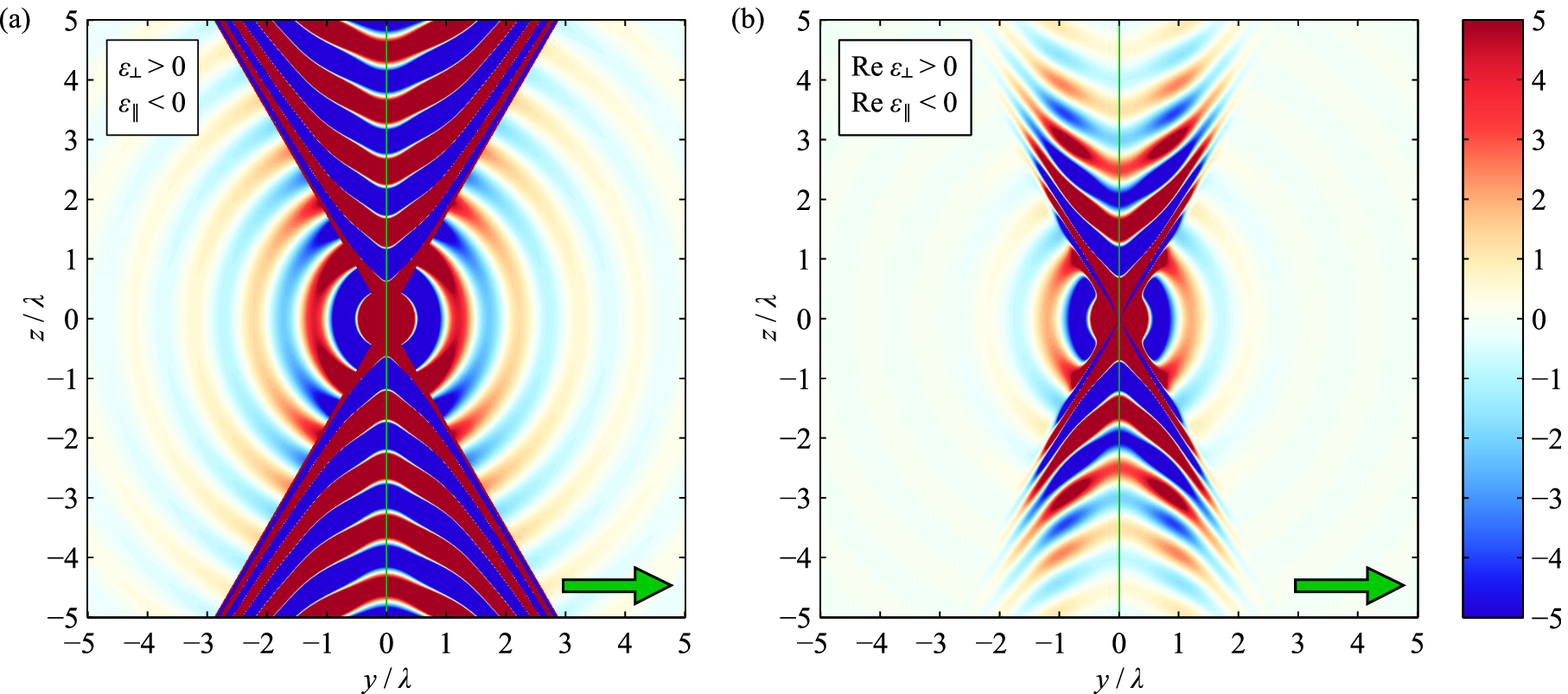}
\end{center}
\caption{Dipole field $\lambda^3 E_y(0, y, z)$ in hyperbolic medium with $\epr = 1$, $\epl = -3$ (a) and with $\epr = 1 + 0.2i$, $\epl = -3 + 0.2i$ (with losses) (b). Dipole moment is orthogonal to the axis of anisotropy~\zz.}\label{fig:Eort1}
\end{figure*}
\begin{figure*}[t]
\begin{center}\includegraphics[width=\textwidth]{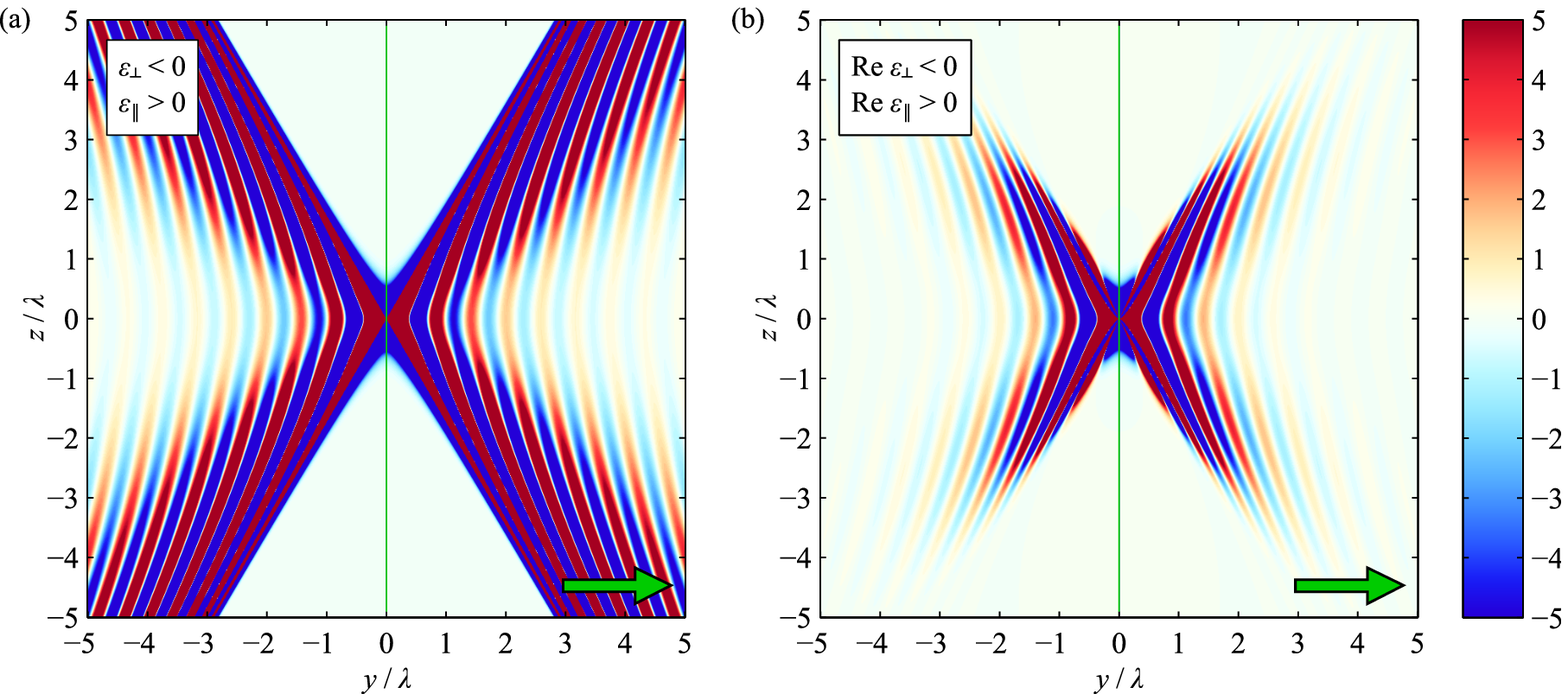}
 \end{center}
\caption{Dipole field $\lambda^3 E_y(0, y, z)$ in hyperbolic medium with $\epr = -1$, $\epl = 3$ (a) and with $\epr = -1 + 0.2i$, $\epl = 3 + 0.2i$ (with losses) (b). Dipole moment is orthogonal to the axis of anisotropy~\zz.}\label{fig:Eort2}
\end{figure*}
\subsection{Dipole parallel to the axis of anisotropy}
Here we consider the case $\vec p = p\zz$. Taking into account that
\begin{gather*}
\eps\cdot\zz=\epr\zz,\quad (\epsn\rr)\cdot\zz = \epr\rr\cdot\zz,\\
(\rr\times\zz)\cdot\zz = 0,\quad (\vec{\hat I} - \zz\otimes\zz)\cdot\zz = 0,
\end{gather*}
we obtain from Eqs.~\eqref{eq:FinalG}, \eqref{eq:SimpleE} the following result for the regular part of the electric field:
\begin{align}
\left[\vec E(\rr)\right]_{\rm reg}&=pk^2\sepr\frac{\eikre}{r_e}\left(1-\frac{1}{ikr_e}-\frac{1}{k^2 r_e^2}\right)\zz\\\nonumber
&-pk^2\sepr\frac{\eikre}{r_e^3}\left(1-\frac{3}{ikr_e}-\frac{3}{k^2r_e^2}\right)(\rr\cdot\zz)(\epsn\rr)\label{eq:AlongE}
\end{align}
Cartesian representation of this expression is presented in \ref{sec:DecartAlong}. We see that the axial dipole emits only TM-polarized (extraordinary) waves.
Calculated cross-section of the electric field in the $yz$ plane is presented in Fig.~\ref{fig:Epar1}, Fig.~\ref{fig:Epar2} for two both cases of hyperbolic medium, illustrated in Fig.~\ref{fig:cones}:
$\Re \epl<0, \Re\epr>0$ and  $\Re \epl>0, \Re\epr<0$, respectively. Electric field pattern has a distinct cone-like shape: the
waves are emitted only within the polar angles $\theta$, satisfying $\epl\sin^2\theta + \epr\cos^2\theta>0$.
This radiation pattern, characteristic for hyperbolic medium, survives even when the losses are introduced, see Fig.~\ref{fig:Epar1}b and Fig.~\ref{fig:Epar2}b.
An interesting feature of Eq.~\eqref{eq:AlongE}, revealed in Fig.~\ref{fig:Epar1}, is that
$|E_z(0,0,z)|$ is not zero for axial dipole orientation and decays as $1/|z|$. This means that in the anisotropic medium the field  is nonzero even along the direction of the dipole $\zz$.

\subsection{Dipole orthogonal to the axis of anisotropy}
Here we consider the case $\vec p = p\yy$. Taking into account that
\begin{gather*}
\epsn\cdot\yy = \epl\yy,\quad (\epsn\rr)\cdot\yy=\epl\rr\cdot\yy,
\end{gather*}
we obtain from~\eqref{eq:FinalG}, \eqref{eq:SimpleE} the following result:
\begin{align}
\left[\vec E(\rr)\right]_{\rm reg}&=pk^2\frac{\epl}{\sepr}\frac{\eikre}{r_e}\left(1-\frac{1}{ikr_e}-\frac{1}{k^2 r_e^2}\right)\yy\nonumber\\
&-pk^2\frac{\epl}{\sepr} \frac{\eikre}{r_e^3}\left(1-\frac{3}{ikr_e}-\frac{3}{k^2r_e^2}\right)(\rr\cdot\yy)(\epsn\rr)\nonumber\\
&+pk^2\left(\sepr\frac{\eikro}{r_o}-\frac{\epl}{\sepr}\frac{\eikre}{r_e}\right)\frac{[(\rc)\cdot\yy](\rc)}{(\rc)^2}\nonumber\\
&-ipk\frac{1}{\sepr}\frac{\eikro-\eikre}{(\rc)^2}\left(\yy- \frac{2[(\rc)\cdot\yy](\rc)}{(\rc)^2}\right)\label{eq:OrthoE}.
\end{align}
Cartesian representation of the last expression is presented in \ref{sec:DecartOrth}. For this geometry both TE and TM polarized waves are emitted.
 Equation~\eqref{eq:OrthoE} includes
the denominator $(\rc)^2\equiv x^2+y^2$ which turns to zero at the line $x=y=0$. However, careful analysis of Eq.~\eqref{eq:OrthoE} ensures, that electric field is continuous at this line since diverging TE and TM wave contributions cancel each other.
Calculated emission pattern is presented in Figs.~\ref{fig:Eort1},\ref{fig:Eort2} for
different signs of dielectric constants.
Most interesting results are manifested
for $\Re \epl<0, \Re\epr>0$ (Fig.~\ref{fig:Eort1}), when the electric field is a distinct superposition of the conic patern due to the TM waves emission and elliptic pattern due to TE waves. In the second case $\Re \epl>0, \Re\epr<0$ (Fig.~\ref{fig:Eort2}) the TE waves contribution leads just to the spatial modulation of the conic radiation pattern. Similar to the case of the axial dipole orientation, the far field is present even along the dipole direction.

\begin{figure*}[t]
\begin{center}
\includegraphics[width=\textwidth]{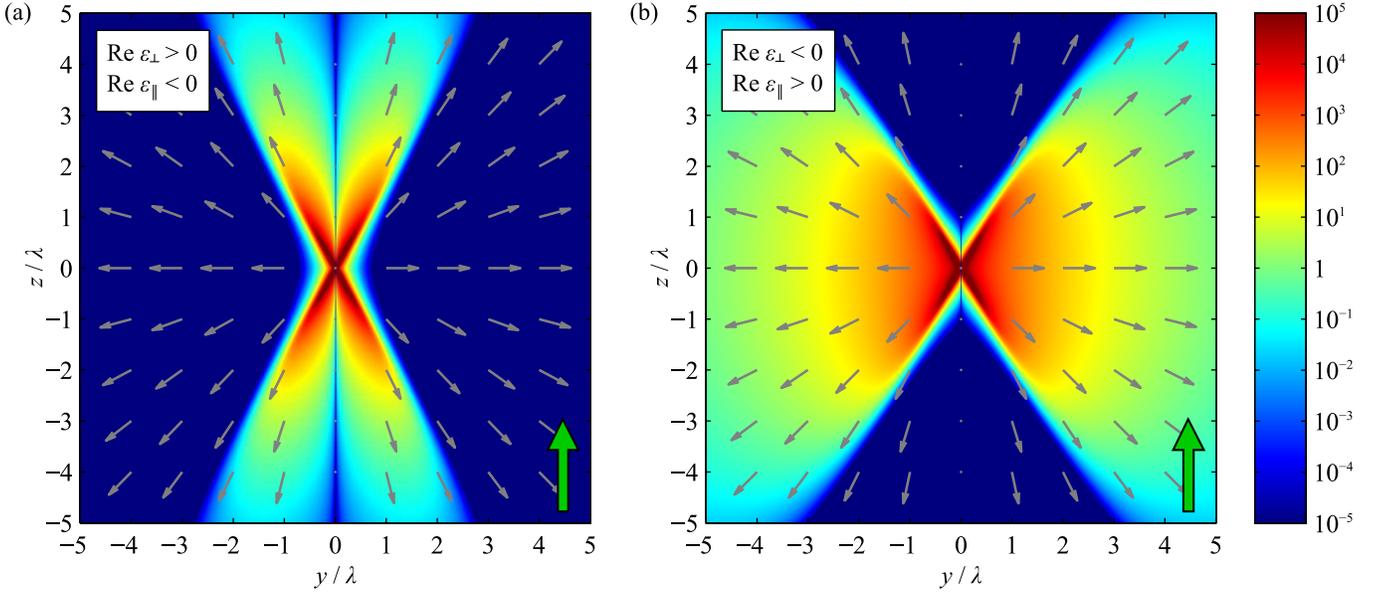}
\end{center}
\caption{Poynting vector $4\pi/(c\lambda^6) \vec S(0, y, z)$ in hyperbolic medium with $\epr = 1+0.2i$, $\epl = -3+0.2i$ (a) and with $\epr = -1 + 0.2i$, $\epl = 3 + 0.2i$ (b). Dipole moment is parallel to the axis of anisotropy~\zz.}
\label{fig:Poynting_par}
\end{figure*}
\begin{figure*}[t]
\begin{center}
\includegraphics[width=\textwidth]{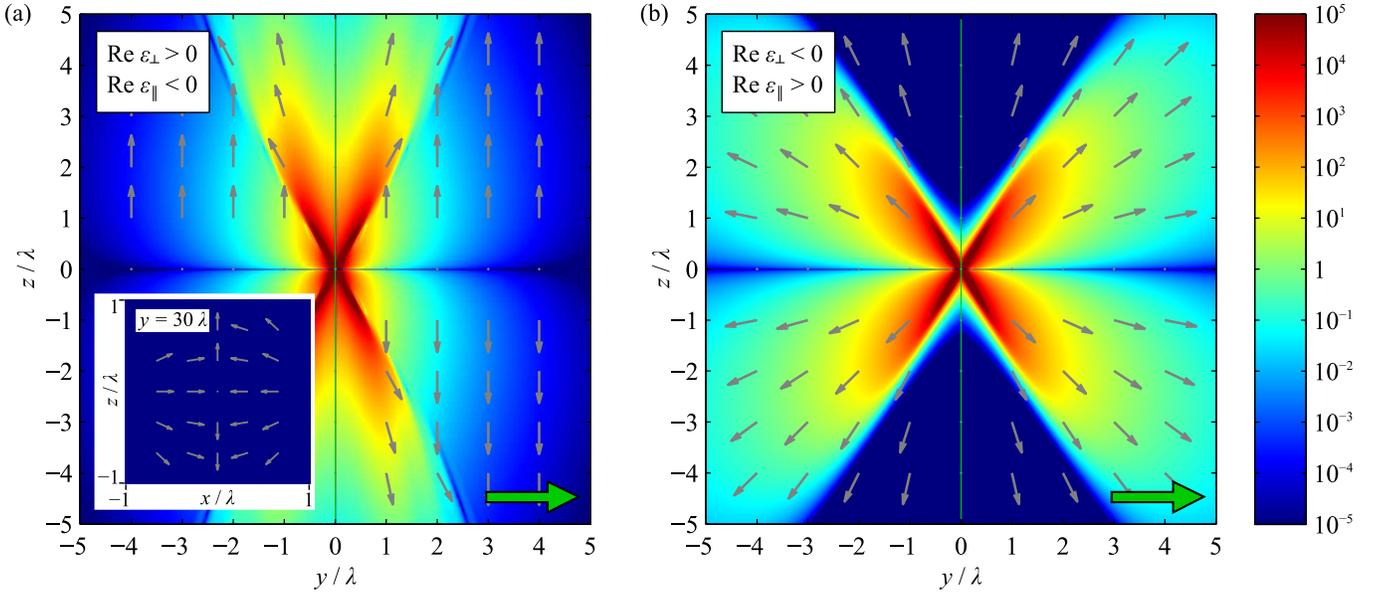}
\end{center}
\caption{Poynting vector $4\pi/(c\lambda^6) \vec S(0, y, z)$ in hyperbolic medium with $\epr = 1+0.2i$, $\epl = -3+0.2i$ (a) and with $\epr = -1 + 0.2i$, $\epl = 3 + 0.2i$ (b). Dipole moment is orthogonal to the axis of anisotropy~\zz.
Inset in panel (a) illustrates the Poynting vector distribution in the $xz$ plane for $y=30\lambda$.
}\label{fig:Poynting_ort}
\end{figure*}

Interesting results are also revealed in the Poynting vector distribution, shown in Figs.~\ref{fig:Poynting_par},\ref{fig:Poynting_ort}.  The Poynting vector is found as
$\bm S=c/(4\pi)\Re \vec E\times\vec H^*$, explicit expressions for magnetic field $\vec H$ are presented in \ref{sec:DecartAlong},\ref{sec:DecartOrth}. The Poynting vector pattern inherits the conical shape from the electric field. Its largest values are  achieved at the conical surface $r_e^2=0$. Interesting features are observed for the Poynting vector distribution in case of perpendicular dipole orientation and $\Re \epl<0$, illustrated on Fig.~\ref{fig:Poynting_ort}a. In the region outside of the cone, $r_e^2<0$, the Poynting vector in the $yz$ plane is directed almost along the $z$ direction. The resulting pattern looks as if there is a line source located at $y=x=0$, which violates the energy conservation condition.  However, this first impression is wrong. Close inspection of the Poynting vector distribution reveals that the line $y=x=0$ is a saddle point of the Poynting vector: the energy enters the line along $x$ direction and leaves along $z$ one. This behavior is illustrated in the $xz$ cross-section of the Poynting vector distribution, shown in the inset of Fig.~\ref{fig:Poynting_ort}a.

One can also prove this result analytically.
 Neglecting the evanescent terms $\propto \exp(-k|r_e|)$ we find  following expression for the Poynting vector for vanishing losses
\begin{align}
 \vec S=&\frac{c}{4\pi r^{3/2}(x^2+y^2)^2\label{eq:S}
\sepr}\\\times\Bigl\{\nonumber
&\xx [k^2\epr x^3(x^2+y^2)+x(x^2-y^2)]\\+\nonumber
&\yy[k^2\epr x^2y(x^2+y^2)+2x^2y]\\+
&\zz [k^2 \epr x^2z (x^2+y^2)+z(r^2+y^2)]\Bigr\}\:.\nonumber
\end{align}
Direct differentiation of Eq.~\eqref{eq:S} demonstrates that $\nabla \cdot \vec S=0$, i.e., energy conservation law is fulfilled. In the limit $|x|,|z|\ll y,1/k$ we find $\vec S\propto (z\zz-x\xx)/|y^4|$, which confirms existence of the saddle point in the Poynting vector pattern.


\section{Conclusions}\label{sec:Concl}

We have presented a general theory of the dipole emission in homogeneous hyperbolic media. Using both Fourier space approach and electromagnetic 
scaling, we have obtained a general expression for the electromagnetic Green function, and demonstrated that the emission pattern is highly anisotropic. 
For dipole orientation parallel to the symmetry axis, only TM-polarized waves are excited and the emission pattern has a cone-like shape with propagating 
waves present only within  the cone $\epl\sin^2\theta+\epr\cos^2\theta>0$, where $\theta$ is the polar angle. 
In case of the perpendicular orientation, the electric field is given by a sum of TE-polarized and TM-polarized contributions, so the waves can propagate 
also outside the cone.  We have revealed the importance of a singular term in the Green function, and have demonstrated that it is of crucial importance 
for calculation of the radiative rate of light source embedded in a hyperbolic medium. In the conventional case, the singular term proportional to the 
$\delta$-function  is usually neglected, and it does not contribute to the Purcell factor. However, in hyperbolic media this singular term is complex 
even for vanishing losses, and it determines the diverging radiative decay rate.

\section{Acknowledgements}

This work has been supported by the Ministry of Education and Science of Russian Federation, the Dynasty Foundation,
Russian Foundation for Basic Research, European project POLAPHEN, EPSRC~(UK),  and the Australian Research Council.
The authors acknowledge useful discussions with S.I.~Maslovski, I.Yu.~Popov, and C.R.~Simovski.
 
\appendix

\section{Cartesian representation of the field of dipole parallel to the anisotropy axis }\label{sec:DecartAlong}
Cartesian components of the electric field Eq.~\eqref{eq:AlongE}
of the dipole oriented along the anisotropy axis $z$
 read
\begin{align}
E_x&=-pk^2\epl\sepr\frac{\eikre}{r_e^3} \left(1-\frac{3}{ikr_e}-\frac{3}{k^2r_e^2}\right)xz,\nonumber\\
E_y&=-pk^2\epl\sepr \frac{\eikre}{r_e^3} \left(1-\frac{3}{ikr_e}-\frac{3}{k^2r_e^2}\right)yz,\label{eq:Decarteq:AlongE}\\
E_z&=\phantom{-}pk^2\epl\sepr\frac{\eikre}{r_e^3}\nonumber\\
&\times\left[x^2+y^2-\left(\frac{1}{ikr_e} + \frac{1}{k^2r_e^2}\right)\left(x^2+y^2-\frac{2\epr}{\epl}z^2\right)\right],\nonumber
\end{align}
where $r_e=\sqrt{\epl(x^2+y^2)+\epr z^2}$.
Similar result has been obtained in~Ref.~\cite{Savchenko2005}.
Magnetic field can be found as $\vec H=-(i/k) \nabla\times\vec E$:
\begin{align}
H_x&=\phantom{-}pk^2\epl\sepr\frac{\eikre}{r_e^2}\left(1-\frac{1}{ikr_e}\right)y,\nonumber\\
H_y&=-pk^2\epl\sepr \frac{\eikre}{r_e^2}\left(1-\frac{1}{ikr_e}\right)x,\\
H_z&=0.\nonumber
\end{align}

\section{Cartesian presentation of the field of dipole orthogonal to the anisotropy axis }\label{sec:DecartOrth}
Cartesian components of Eq.~\eqref{eq:OrthoE} for the dipole oriented along $y$ axis, perpendicular to the anisotropy axis $z$, read:\cite{Savchenko2005}
\begin{align}
E_x &=-pk^2\frac{\epl^2}{\sepr}\frac{\eikre}{r_e^3}\left(1-\frac{3}{ikr_e}-\frac{3}{k^2r_e^2}\right)xy\nonumber\\
&\phantom{=}-pk^2\left(\sepr\frac{\eikro}{r_o}-\frac{\epl}{\sepr}\frac{\eikre}{r_e}\right)\frac{xy}{x^2+y^2}\nonumber\\
&\phantom{=}-ipk\frac{1}{\sepr}\left(\eikro - \eikre\right)\frac{2xy}{(x^2+y^2)^2}\nonumber\\
E_y &=\phantom{-}pk^2\frac{\epl^2}{\sepr}\frac{\eikre}{r_e^3} \label{eq:Decarteq:OrthoE}\\
&\times \left[x^2+\frac{\epr}{\epl}z^2-\left(\frac{1}{ikr_e}+\frac{1}{k^2r_e^2}\right)\left(x^2+\frac{\epr}{\epl}z^2-2y^2\right)\right]\nonumber\\
&\phantom{=}+pk^2\left(\sepr\frac{\eikro}{r_o}-\frac{\epl}{\sepr}\frac{\eikre}{r_e}\right)\frac{x^2}{x^2+y^2}\nonumber\\
&\phantom{=}+ipk\frac{1}{\sepr}\left(\eikro - \eikre\right)\frac{x^2-y^2}{(x^2+y^2)^2},\nonumber\\
E_z &=-pk^2\epl\sepr\frac{\eikre}{r_e^3}\left(1-\frac{3}{ikr_e}-\frac{3}{k^2r_e^2}\right)yz,\nonumber
\end{align}
where
\begin{gather*}
r_e=\sqrt{\epl(x^2+y^2)+\epr z^2},\\
r_o=\sqrt{\epr(x^2+y^2+z^2)}.
\end{gather*}
Magnetic field reads
\begin{align}
H_x&=-pk^2\epl\sepr\frac{\eikre}{r_e^2}\left(1-\frac{1}{ikr_e}\right)\frac{y^2z}{x^2+y^2}\nonumber\\
&\phantom{ =}-pk^2\epr\sepr\frac{\eikro}{r_o^2}\left(1-\frac{1}{ikr_o}\right)\frac{x^2z}{x^2+y^2}\nonumber\\
&\phantom{=}-ipk\sepr\left(\frac{\eikro}{r_o}-\frac{\eikre}{r_e}\right)\frac{(x^2-y^2)z}{(x^2+y^2)^2},\nonumber\\
H_y&=\phantom{-}pk^2\epl\sepr\frac{\eikre}{r_e^2}\left(1-\frac{1}{ikr_e}\right)\frac{xyz}{x^2+y^2}\\
&\phantom{=}-pk^2\epr\sepr\frac{\eikro}{r_o}\left(1-\frac{1}{ikr_o}\right)\frac{xyz}{x^2+y^2}\nonumber\\
&\phantom{=}-ipk\sepr\left(\frac{\eikro}{r_o}-\frac{\eikre}{r_e}\right)\frac{2xyz}{(x^2+y^2)^2},\nonumber\\
H_z&=\phantom{+}pk^2\epr\sepr\frac{\eikro}{r_o^2}\left(1-\frac{1}{ikr_o}\right)x.\nonumber
\end{align}

\begin{thebibliography}{44}
\expandafter\ifx\csname natexlab\endcsname\relax\def\natexlab#1{#1}\fi
\expandafter\ifx\csname bibnamefont\endcsname\relax
  \def\bibnamefont#1{#1}\fi
\expandafter\ifx\csname bibfnamefont\endcsname\relax
  \def\bibfnamefont#1{#1}\fi
\expandafter\ifx\csname citenamefont\endcsname\relax
  \def\citenamefont#1{#1}\fi
\expandafter\ifx\csname url\endcsname\relax
  \def\url#1{\texttt{#1}}\fi
\expandafter\ifx\csname urlprefix\endcsname\relax\def\urlprefix{URL }\fi
\providecommand{\bibinfo}[2]{#2}
\providecommand{\eprint}[2][]{\url{#2}}

\bibitem[{\citenamefont{Felsen and Marcuvitz}(1972)}]{Felsen1972}
\bibinfo{author}{\bibfnamefont{L.~B.} \bibnamefont{Felsen}} \bibnamefont{and}
  \bibinfo{author}{\bibfnamefont{N.}~\bibnamefont{Marcuvitz}},
  \emph{\bibinfo{title}{Radiation and scattering of waves}}
  (\bibinfo{publisher}{Prentice-Hall Englewood Cliffs, N.J.,},
  \bibinfo{year}{1972}), ISBN \bibinfo{isbn}{0137503644}.

\bibitem[{\citenamefont{Smith and Schurig}(2003)}]{Smith2003}
\bibinfo{author}{\bibfnamefont{D.~R.} \bibnamefont{Smith}} \bibnamefont{and}
  \bibinfo{author}{\bibfnamefont{D.}~\bibnamefont{Schurig}},
  \bibinfo{journal}{Phys. Rev. Lett.} \textbf{\bibinfo{volume}{90}},
  \bibinfo{pages}{077405} (\bibinfo{year}{2003}).

\bibitem[{\citenamefont{Smith et~al.}(2004)\citenamefont{Smith, Kolinko, and
  Schurig}}]{Smith2004}
\bibinfo{author}{\bibfnamefont{D.~R.} \bibnamefont{Smith}},
  \bibinfo{author}{\bibfnamefont{P.}~\bibnamefont{Kolinko}}, \bibnamefont{and}
  \bibinfo{author}{\bibfnamefont{D.}~\bibnamefont{Schurig}},
  \bibinfo{journal}{J. Opt. Soc. Am. B} \textbf{\bibinfo{volume}{21}},
  \bibinfo{pages}{1032} (\bibinfo{year}{2004}).

\bibitem[{\citenamefont{Cai et~al.}(2008)\citenamefont{Cai, Chettiar,
  Kildishev, and Shalaev}}]{Cai2008}
\bibinfo{author}{\bibfnamefont{W.}~\bibnamefont{Cai}},
  \bibinfo{author}{\bibfnamefont{U.~K.} \bibnamefont{Chettiar}},
  \bibinfo{author}{\bibfnamefont{A.~V.} \bibnamefont{Kildishev}},
  \bibnamefont{and} \bibinfo{author}{\bibfnamefont{V.~M.}
  \bibnamefont{Shalaev}}, \bibinfo{journal}{Opt. Express}
  \textbf{\bibinfo{volume}{16}}, \bibinfo{pages}{5444} (\bibinfo{year}{2008}).

\bibitem[{\citenamefont{Krishnamoorthy
  et~al.}(2012)\citenamefont{Krishnamoorthy, Jacob, Narimanov, Kretzschmar, and
  Menon}}]{Krishnamoorthy2012}
\bibinfo{author}{\bibfnamefont{H.~N.~S.} \bibnamefont{Krishnamoorthy}},
  \bibinfo{author}{\bibfnamefont{Z.}~\bibnamefont{Jacob}},
  \bibinfo{author}{\bibfnamefont{E.}~\bibnamefont{Narimanov}},
  \bibinfo{author}{\bibfnamefont{I.}~\bibnamefont{Kretzschmar}},
  \bibnamefont{and} \bibinfo{author}{\bibfnamefont{V.~M.} \bibnamefont{Menon}},
  \bibinfo{journal}{Science} \textbf{\bibinfo{volume}{336}},
  \bibinfo{pages}{205} (\bibinfo{year}{2012}).

\bibitem[{\citenamefont{{Jacob} and {Shalaev}}(2011)}]{shalaev2011}
\bibinfo{author}{\bibfnamefont{Z.}~\bibnamefont{{Jacob}}} \bibnamefont{and}
  \bibinfo{author}{\bibfnamefont{V.~M.} \bibnamefont{{Shalaev}}},
  \bibinfo{journal}{Science} \textbf{\bibinfo{volume}{334}},
  \bibinfo{pages}{463} (\bibinfo{year}{2011}).

\bibitem[{\citenamefont{Purcell}(1946)}]{Purcell}
\bibinfo{author}{\bibfnamefont{E.~M.} \bibnamefont{Purcell}},
  \bibinfo{journal}{Phys. Rev.} \textbf{\bibinfo{volume}{69}},
  \bibinfo{pages}{681} (\bibinfo{year}{1946}).

\bibitem[{\citenamefont{Poddubny et~al.}(2011)\citenamefont{Poddubny, Belov,
  and Kivshar}}]{Poddubny2011}
\bibinfo{author}{\bibfnamefont{A.~N.} \bibnamefont{Poddubny}},
  \bibinfo{author}{\bibfnamefont{P.~A.} \bibnamefont{Belov}}, \bibnamefont{and}
  \bibinfo{author}{\bibfnamefont{Y.~S.} \bibnamefont{Kivshar}},
  \bibinfo{journal}{Phys. Rev. A} \textbf{\bibinfo{volume}{84}},
  \bibinfo{pages}{023807} (\bibinfo{year}{2011}).

\bibitem[{\citenamefont{Kidwai et~al.}(2011)\citenamefont{Kidwai, Zhukovsky,
  and Sipe}}]{sipe2011}
\bibinfo{author}{\bibfnamefont{O.}~\bibnamefont{Kidwai}},
  \bibinfo{author}{\bibfnamefont{S.~V.} \bibnamefont{Zhukovsky}},
  \bibnamefont{and} \bibinfo{author}{\bibfnamefont{J.~E.} \bibnamefont{Sipe}},
  \bibinfo{journal}{Opt. Lett.} \textbf{\bibinfo{volume}{36}},
  \bibinfo{pages}{2530} (\bibinfo{year}{2011}).

\bibitem[{\citenamefont{Iorsh et~al.}(2012)\citenamefont{Iorsh, Poddubny,
  Orlov, Belov, and Kivshar}}]{Iorsh2012}
\bibinfo{author}{\bibfnamefont{I.}~\bibnamefont{Iorsh}},
  \bibinfo{author}{\bibfnamefont{A.}~\bibnamefont{Poddubny}},
  \bibinfo{author}{\bibfnamefont{A.}~\bibnamefont{Orlov}},
  \bibinfo{author}{\bibfnamefont{P.}~\bibnamefont{Belov}}, \bibnamefont{and}
  \bibinfo{author}{\bibfnamefont{Y.~S.} \bibnamefont{Kivshar}},
  \bibinfo{journal}{Phys. Lett. A} \textbf{\bibinfo{volume}{376}},
  \bibinfo{pages}{185 } (\bibinfo{year}{2012}).

\bibitem[{\citenamefont{{Yan} et~al.}(2012)\citenamefont{{Yan}, {Wubs}, and
  {Asger Mortensen}}}]{Yan2012arXiv}
\bibinfo{author}{\bibfnamefont{W.}~\bibnamefont{{Yan}}},
  \bibinfo{author}{\bibfnamefont{M.}~\bibnamefont{{Wubs}}}, \bibnamefont{and}
  \bibinfo{author}{\bibfnamefont{N.}~\bibnamefont{{Asger Mortensen}}},
  \bibinfo{journal}{ArXiv e-prints}  (\bibinfo{year}{2012}),
  \eprint{1204.5413}.

\bibitem[{\citenamefont{Kim et~al.}(2012)\citenamefont{Kim, Drachev, Jacob,
  Naik, Boltasseva, Narimanov, and Shalaev}}]{Kim2012}
\bibinfo{author}{\bibfnamefont{J.}~\bibnamefont{Kim}},
  \bibinfo{author}{\bibfnamefont{V.~P.} \bibnamefont{Drachev}},
  \bibinfo{author}{\bibfnamefont{Z.}~\bibnamefont{Jacob}},
  \bibinfo{author}{\bibfnamefont{G.~V.} \bibnamefont{Naik}},
  \bibinfo{author}{\bibfnamefont{A.}~\bibnamefont{Boltasseva}},
  \bibinfo{author}{\bibfnamefont{E.~E.} \bibnamefont{Narimanov}},
  \bibnamefont{and} \bibinfo{author}{\bibfnamefont{V.~M.}
  \bibnamefont{Shalaev}}, \bibinfo{journal}{Opt. Express}
  \textbf{\bibinfo{volume}{20}}, \bibinfo{pages}{8100} (\bibinfo{year}{2012}).

\bibitem[{\citenamefont{Fisher and Gould}(1969)}]{Fisher1969}
\bibinfo{author}{\bibfnamefont{R.~K.} \bibnamefont{Fisher}} \bibnamefont{and}
  \bibinfo{author}{\bibfnamefont{R.~W.} \bibnamefont{Gould}},
  \bibinfo{journal}{Phys. Rev. Lett.} \textbf{\bibinfo{volume}{22}},
  \bibinfo{pages}{1093} (\bibinfo{year}{1969}).

\bibitem[{\citenamefont{Zhang et~al.}(2011)\citenamefont{Zhang, Xiong, Bartal,
  Yin, and Zhang}}]{Zhang2011}
\bibinfo{author}{\bibfnamefont{S.}~\bibnamefont{Zhang}},
  \bibinfo{author}{\bibfnamefont{Y.}~\bibnamefont{Xiong}},
  \bibinfo{author}{\bibfnamefont{G.}~\bibnamefont{Bartal}},
  \bibinfo{author}{\bibfnamefont{X.}~\bibnamefont{Yin}}, \bibnamefont{and}
  \bibinfo{author}{\bibfnamefont{X.}~\bibnamefont{Zhang}},
  \bibinfo{journal}{Phys. Rev. Lett.} \textbf{\bibinfo{volume}{106}},
  \bibinfo{pages}{243901} (\bibinfo{year}{2011}).

\bibitem[{\citenamefont{Yao et~al.}(2008)\citenamefont{Yao, Liu, Liu, Wang,
  Sun, Bartal, Stacy, and Zhang}}]{Yao2008}
\bibinfo{author}{\bibfnamefont{J.}~\bibnamefont{Yao}},
  \bibinfo{author}{\bibfnamefont{Z.}~\bibnamefont{Liu}},
  \bibinfo{author}{\bibfnamefont{Y.}~\bibnamefont{Liu}},
  \bibinfo{author}{\bibfnamefont{Y.}~\bibnamefont{Wang}},
  \bibinfo{author}{\bibfnamefont{C.}~\bibnamefont{Sun}},
  \bibinfo{author}{\bibfnamefont{G.}~\bibnamefont{Bartal}},
  \bibinfo{author}{\bibfnamefont{A.~M.} \bibnamefont{Stacy}}, \bibnamefont{and}
  \bibinfo{author}{\bibfnamefont{X.}~\bibnamefont{Zhang}},
  \bibinfo{journal}{Science} \textbf{\bibinfo{volume}{321}},
  \bibinfo{pages}{930} (\bibinfo{year}{2008}).

\bibitem[{\citenamefont{Noginov et~al.}(2009)\citenamefont{Noginov, Barnakov,
  Zhu, Tumkur, Li, and Narimanov}}]{narimanov2009b}
\bibinfo{author}{\bibfnamefont{M.~A.} \bibnamefont{Noginov}},
  \bibinfo{author}{\bibfnamefont{Y.~A.} \bibnamefont{Barnakov}},
  \bibinfo{author}{\bibfnamefont{G.}~\bibnamefont{Zhu}},
  \bibinfo{author}{\bibfnamefont{T.}~\bibnamefont{Tumkur}},
  \bibinfo{author}{\bibfnamefont{H.}~\bibnamefont{Li}}, \bibnamefont{and}
  \bibinfo{author}{\bibfnamefont{E.~E.} \bibnamefont{Narimanov}},
  \bibinfo{journal}{Appl. Phys. Lett.} \textbf{\bibinfo{volume}{94}},
  \bibinfo{eid}{151105} (pages~\bibinfo{numpages}{3}) (\bibinfo{year}{2009}).

\bibitem[{\citenamefont{{Alekseyev} et~al.}(2010)\citenamefont{{Alekseyev},
  {Narimanov}, {Tumkur}, {Li}, {Barnakov}, and {Noginov}}}]{narimanov2010b}
\bibinfo{author}{\bibfnamefont{L.~V.} \bibnamefont{{Alekseyev}}},
  \bibinfo{author}{\bibfnamefont{E.~E.} \bibnamefont{{Narimanov}}},
  \bibinfo{author}{\bibfnamefont{T.}~\bibnamefont{{Tumkur}}},
  \bibinfo{author}{\bibfnamefont{H.}~\bibnamefont{{Li}}},
  \bibinfo{author}{\bibfnamefont{Y.~A.} \bibnamefont{{Barnakov}}},
  \bibnamefont{and} \bibinfo{author}{\bibfnamefont{M.~A.}
  \bibnamefont{{Noginov}}}, \bibinfo{journal}{Appl. Phys. Lett.}
  \textbf{\bibinfo{volume}{97}}, \bibinfo{pages}{131107}
  (\bibinfo{year}{2010}).

\bibitem[{\citenamefont{Cortes et~al.}(2012)\citenamefont{Cortes, Newman,
  Molesky, and Jacob}}]{Cortes2012arXiv}
\bibinfo{author}{\bibfnamefont{C.~L.} \bibnamefont{Cortes}},
  \bibinfo{author}{\bibfnamefont{W.}~\bibnamefont{Newman}},
  \bibinfo{author}{\bibfnamefont{S.}~\bibnamefont{Molesky}}, \bibnamefont{and}
  \bibinfo{author}{\bibfnamefont{Z.}~\bibnamefont{Jacob}},
  \bibinfo{journal}{Journal of Optics} \textbf{\bibinfo{volume}{14}},
  \bibinfo{pages}{063001} (\bibinfo{year}{2012}).

\bibitem[{\citenamefont{Orlov et~al.}(2011)\citenamefont{Orlov, Voroshilov,
  Belov, and Kivshar}}]{orlov2011}
\bibinfo{author}{\bibfnamefont{A.~A.} \bibnamefont{Orlov}},
  \bibinfo{author}{\bibfnamefont{P.~M.} \bibnamefont{Voroshilov}},
  \bibinfo{author}{\bibfnamefont{P.~A.} \bibnamefont{Belov}}, \bibnamefont{and}
  \bibinfo{author}{\bibfnamefont{Y.~S.} \bibnamefont{Kivshar}},
  \bibinfo{journal}{Phys. Rev. B} \textbf{\bibinfo{volume}{84}},
  \bibinfo{pages}{045424} (\bibinfo{year}{2011}).

\bibitem[{\citenamefont{Chebykin et~al.}(2011)\citenamefont{Chebykin, Orlov,
  Vozianova, Maslovski, Kivshar, and Belov}}]{chebykin2011}
\bibinfo{author}{\bibfnamefont{A.~V.} \bibnamefont{Chebykin}},
  \bibinfo{author}{\bibfnamefont{A.~A.} \bibnamefont{Orlov}},
  \bibinfo{author}{\bibfnamefont{A.~V.} \bibnamefont{Vozianova}},
  \bibinfo{author}{\bibfnamefont{S.~I.} \bibnamefont{Maslovski}},
  \bibinfo{author}{\bibfnamefont{Y.~S.} \bibnamefont{Kivshar}},
  \bibnamefont{and} \bibinfo{author}{\bibfnamefont{P.~A.} \bibnamefont{Belov}},
  \bibinfo{journal}{Phys. Rev. B} \textbf{\bibinfo{volume}{84}},
  \bibinfo{pages}{115438} (\bibinfo{year}{2011}).

\bibitem[{\citenamefont{Wurtz et~al.}(2008)\citenamefont{Wurtz, Dickson,
  O'Connor, Atkinson, Hendren, Evans, Pollard, and Zayats}}]{Wurtz2008}
\bibinfo{author}{\bibfnamefont{G.~A.} \bibnamefont{Wurtz}},
  \bibinfo{author}{\bibfnamefont{W.}~\bibnamefont{Dickson}},
  \bibinfo{author}{\bibfnamefont{D.}~\bibnamefont{O'Connor}},
  \bibinfo{author}{\bibfnamefont{R.}~\bibnamefont{Atkinson}},
  \bibinfo{author}{\bibfnamefont{W.}~\bibnamefont{Hendren}},
  \bibinfo{author}{\bibfnamefont{P.}~\bibnamefont{Evans}},
  \bibinfo{author}{\bibfnamefont{R.}~\bibnamefont{Pollard}}, \bibnamefont{and}
  \bibinfo{author}{\bibfnamefont{A.~V.} \bibnamefont{Zayats}},
  \bibinfo{journal}{Opt. Express} \textbf{\bibinfo{volume}{16}},
  \bibinfo{pages}{7460} (\bibinfo{year}{2008}).

\bibitem[{\citenamefont{Simovski et~al.}(2012)\citenamefont{Simovski, Belov,
  Atrashchenko, and Kivshar}}]{simovski2012}
\bibinfo{author}{\bibfnamefont{C.~R.} \bibnamefont{Simovski}},
  \bibinfo{author}{\bibfnamefont{P.~A.} \bibnamefont{Belov}},
  \bibinfo{author}{\bibfnamefont{A.~V.} \bibnamefont{Atrashchenko}},
  \bibnamefont{and} \bibinfo{author}{\bibfnamefont{Y.~S.}
  \bibnamefont{Kivshar}}, \bibinfo{journal}{Adv. Materials}
  (\bibinfo{year}{2012}), \bibinfo{note}{in press}.

\bibitem[{\citenamefont{Sun et~al.}(2011)\citenamefont{Sun, Zhou, Li, and
  Kang}}]{Sun2011}
\bibinfo{author}{\bibfnamefont{J.}~\bibnamefont{Sun}},
  \bibinfo{author}{\bibfnamefont{J.}~\bibnamefont{Zhou}},
  \bibinfo{author}{\bibfnamefont{B.}~\bibnamefont{Li}}, \bibnamefont{and}
  \bibinfo{author}{\bibfnamefont{F.}~\bibnamefont{Kang}},
  \bibinfo{journal}{Appl. Phys. Lett.} \textbf{\bibinfo{volume}{98}},
  \bibinfo{eid}{101901} (\bibinfo{year}{2011}).

\bibitem[{\citenamefont{Vogel and Welsch}(2006)}]{Welsch2006}
\bibinfo{author}{\bibfnamefont{W.}~\bibnamefont{Vogel}} \bibnamefont{and}
  \bibinfo{author}{\bibfnamefont{D.-G.} \bibnamefont{Welsch}},
  \emph{\bibinfo{title}{Quantum Optics}} (\bibinfo{publisher}{Wiley},
  \bibinfo{address}{Weinheim}, \bibinfo{year}{2006}).

\bibitem[{\citenamefont{Clemmow}(1963{\natexlab{a}})}]{Clemmow1963}
\bibinfo{author}{\bibfnamefont{P.~C.} \bibnamefont{Clemmow}},
  \bibinfo{journal}{Proc. Inst. Elect. Eng.} \textbf{\bibinfo{volume}{110}},
  \bibinfo{pages}{107 } (\bibinfo{year}{1963}{\natexlab{a}}).

\bibitem[{\citenamefont{Chen}(1983)}]{Chen1983}
\bibinfo{author}{\bibfnamefont{H.~C.} \bibnamefont{Chen}},
  \emph{\bibinfo{title}{Theory of electromagnetic waves: a coordinate-free
  approach}}, McGraw-Hill series in electrical engineering
  (\bibinfo{publisher}{McGraw-Hill Book Co.}, \bibinfo{year}{1983}).

\bibitem[{\citenamefont{Weiglhofer}(1988)}]{weiglhofer:1095}
\bibinfo{author}{\bibfnamefont{W.}~\bibnamefont{Weiglhofer}},
  \bibinfo{journal}{Am. J. Phys.} \textbf{\bibinfo{volume}{56}},
  \bibinfo{pages}{1095} (\bibinfo{year}{1988}).

\bibitem[{\citenamefont{Weiglhofer}(1989)}]{Weiglhofer1989}
\bibinfo{author}{\bibfnamefont{W.}~\bibnamefont{Weiglhofer}},
  \bibinfo{journal}{Am. J. Phys.} \textbf{\bibinfo{volume}{57}},
  \bibinfo{pages}{455} (\bibinfo{year}{1989}).

\bibitem[{\citenamefont{Weiglhofer}(1990)}]{Weiglhofer1990}
\bibinfo{author}{\bibfnamefont{W.}~\bibnamefont{Weiglhofer}},
  \bibinfo{journal}{IEE Proc. H} \textbf{\bibinfo{volume}{137}},
  \bibinfo{pages}{5 } (\bibinfo{year}{1990}).

\bibitem[{\citenamefont{Lindell}(1992)}]{Lindell1992}
\bibinfo{author}{\bibfnamefont{I.~V.} \bibnamefont{Lindell}},
  \emph{\bibinfo{title}{Methods for electromagnetic field analysis}}
  (\bibinfo{publisher}{Clarendon Press; Oxford University Press},
  \bibinfo{address}{Oxford; New York}, \bibinfo{year}{1992}).

\bibitem[{\citenamefont{Cottis et~al.}(1999)\citenamefont{Cottis, Vazouras, and
  Spyrou}}]{cottis1999}
\bibinfo{author}{\bibfnamefont{P.}~\bibnamefont{Cottis}},
  \bibinfo{author}{\bibfnamefont{C.}~\bibnamefont{Vazouras}}, \bibnamefont{and}
  \bibinfo{author}{\bibfnamefont{C.}~\bibnamefont{Spyrou}},
  \bibinfo{journal}{IEEE Trans. Antennas Propag.}
  \textbf{\bibinfo{volume}{47}}, \bibinfo{pages}{195 } (\bibinfo{year}{1999}).

\bibitem[{\citenamefont{Savchenko and Savchenko}(2005)}]{Savchenko2005}
\bibinfo{author}{\bibfnamefont{A.}~\bibnamefont{Savchenko}} \bibnamefont{and}
  \bibinfo{author}{\bibfnamefont{O.}~\bibnamefont{Savchenko}},
  \bibinfo{journal}{Technical Physics} \textbf{\bibinfo{volume}{50}},
  \bibinfo{pages}{1366} (\bibinfo{year}{2005}).

\bibitem[{\citenamefont{Sautbekov~S.}(2008)}]{Sautbekov2008}
\bibinfo{author}{\bibfnamefont{F.~P.} \bibnamefont{Sautbekov~S.},
  \bibfnamefont{Kanymgazieva~I.}}, \bibinfo{journal}{Journal of Applied
  Electromagnetism (JAE)} \textbf{\bibinfo{volume}{10, 2}}, \bibinfo{pages}{43}
  (\bibinfo{year}{2008}).

\bibitem[{\citenamefont{Gelfand and Shilov}(1964)}]{Gelfand}
\bibinfo{author}{\bibfnamefont{I.~M.} \bibnamefont{Gelfand}} \bibnamefont{and}
  \bibinfo{author}{\bibfnamefont{G.~E.} \bibnamefont{Shilov}},
  \emph{\bibinfo{title}{Generalized Functions. Volume I: Properties and
  Operations}} (\bibinfo{publisher}{Academic Press}, \bibinfo{year}{1964}).

\bibitem[{\citenamefont{Yaghjian}(1980)}]{Yaghjian1980}
\bibinfo{author}{\bibfnamefont{A.}~\bibnamefont{Yaghjian}},
  \bibinfo{journal}{Proc. IEEE} \textbf{\bibinfo{volume}{68}},
  \bibinfo{pages}{248 } (\bibinfo{year}{1980}), ISSN \bibinfo{issn}{0018-9219}.

\bibitem[{\citenamefont{Tai and Yaghjian}(1981)}]{Tai1981}
\bibinfo{author}{\bibfnamefont{C.}~\bibnamefont{Tai}} \bibnamefont{and}
  \bibinfo{author}{\bibfnamefont{A.}~\bibnamefont{Yaghjian}},
  \bibinfo{journal}{Proc. IEEE} \textbf{\bibinfo{volume}{69}},
  \bibinfo{pages}{282 } (\bibinfo{year}{1981}).

\bibitem[{\citenamefont{Wang}(1982)}]{Wang1982}
\bibinfo{author}{\bibfnamefont{J.}~\bibnamefont{Wang}}, \bibinfo{journal}{IEEE
  Trans. Antennas Propag.} \textbf{\bibinfo{volume}{30}}, \bibinfo{pages}{463 }
  (\bibinfo{year}{1982}).

\bibitem[{\citenamefont{Franklin}(2010)}]{Franklin2010}
\bibinfo{author}{\bibfnamefont{J.}~\bibnamefont{Franklin}},
  \bibinfo{journal}{Am. J. Phys.} \textbf{\bibinfo{volume}{78}},
  \bibinfo{pages}{1225} (\bibinfo{year}{2010}).

\bibitem[{\citenamefont{Frahm}(1983)}]{Frahm1983}
\bibinfo{author}{\bibfnamefont{C.~P.} \bibnamefont{Frahm}},
  \bibinfo{journal}{Am. J. Phys.} \textbf{\bibinfo{volume}{51}},
  \bibinfo{pages}{826} (\bibinfo{year}{1983}).

\bibitem[{\citenamefont{Novotny and Hecht}(2006)}]{Novotny2006}
\bibinfo{author}{\bibfnamefont{L.}~\bibnamefont{Novotny}} \bibnamefont{and}
  \bibinfo{author}{\bibfnamefont{B.}~\bibnamefont{Hecht}},
  \emph{\bibinfo{title}{Principles of Nano-Optics}}
  (\bibinfo{publisher}{Cambridge University Press}, \bibinfo{address}{New
  York}, \bibinfo{year}{2006}).

\bibitem[{\citenamefont{Ivchenko}(2005)}]{Ivchenko2005}
\bibinfo{author}{\bibfnamefont{E.~L.} \bibnamefont{Ivchenko}},
  \emph{\bibinfo{title}{Optical spectroscopy of semiconductor nanostructures}}
  (\bibinfo{publisher}{Alpha Science International}, \bibinfo{address}{Harrow,
  UK}, \bibinfo{year}{2005}).

\bibitem[{\citenamefont{Clemmow}(1963{\natexlab{b}})}]{Clemmow1963a}
\bibinfo{author}{\bibfnamefont{P.~C.} \bibnamefont{Clemmow}},
  \bibinfo{journal}{Proc. Inst. Elect. Eng.} \textbf{\bibinfo{volume}{110}},
  \bibinfo{pages}{101} (\bibinfo{year}{1963}{\natexlab{b}}).

\bibitem[{\citenamefont{Lindell}(1988)}]{Lindell1988}
\bibinfo{author}{\bibfnamefont{I.}~\bibnamefont{Lindell}},
  \bibinfo{journal}{IEEE Trans. Antennas Propag.}
  \textbf{\bibinfo{volume}{36}}, \bibinfo{pages}{1382 } (\bibinfo{year}{1988}).

\bibitem[{\citenamefont{Lindell}(1990)}]{Lindell1990}
\bibinfo{author}{\bibfnamefont{I.}~\bibnamefont{Lindell}},
  \bibinfo{journal}{IEEE Trans. Antennas Propag.}
  \textbf{\bibinfo{volume}{38}}, \bibinfo{pages}{353 } (\bibinfo{year}{1990}).

\end{thebibliography}

\end{document}